\documentclass[prd,superscriptaddress,floatfix,notitlepage,nofootinbib,reprint]{revtex4-1} 

\usepackage{graphicx}  
\usepackage{dcolumn}   
\usepackage{bm}        
\usepackage{amsmath,amssymb}
\usepackage{aas_macros}
\usepackage{multirow}
\usepackage{color}
\usepackage{verbatim}
\usepackage{url}
\usepackage{hyperref}

\newcommand{\mr}{\mathrm}
\newcommand{\kpa}{k_\parallel}
\newcommand{\kpe}{k_\perp}
\newcommand{\bmk}{\bm{k}}
\newcommand{\bmx}{\bm{x}}
\newcommand{\bea}{\begin{equation}}
\newcommand{\eea}{\end{equation}}

\begin{document}
\widetext

\title{Recovering lost 21 cm radial modes via cosmic tidal reconstruction}

\author{Hong-Ming Zhu}
\email[]{hmzhu@berkeley.edu}
\affiliation{Key Laboratory for Computational Astrophysics, National Astronomical Observatories, Chinese Academy of Sciences, 20A Datun Road, Beijing 100012, China}
\affiliation{University of Chinese Academy of Sciences, Beijing 100049, China}
\affiliation{Berkeley Center for Cosmological Physics and Department of Physics,
University of California, Berkeley, California 94720, USA}

\author{Ue-Li Pen} 
\affiliation{Canadian Institute for Theoretical Astrophysics, University of Toronto, 60 St. George Street, Toronto, Ontario M5S 3H8, Canada}
\affiliation{Dunlap Institute for Astronomy and Astrophysics, University of Toronto, 50 St. George Street, Toronto, Ontario M5S 3H4, Canada}
\affiliation{Canadian Institute for Advanced Research, CIFAR Program in Gravitation and Cosmology, Toronto, Ontario M5G 1Z8, Canada}
\affiliation{Perimeter Institute for Theoretical Physics, 31 Caroline Street North, Waterloo, Ontario N2L 2Y5, Canada}

\author{Yu Yu}
\affiliation{Department of Astronomy, Shanghai Jiao Tong University, 800 Dongchuan Road, Shanghai 200240, China}

\author{Xuelei Chen}
\affiliation{Key Laboratory for Computational Astrophysics,  National Astronomical Observatories, Chinese Academy of Sciences, 20A Datun Road, Beijing 100012, China}
\affiliation{University of Chinese Academy of Sciences, Beijing 100049, China}
\affiliation{Center of High Energy Physics, Peking University, Beijing 100871, China}
\date{\today}

\begin{abstract}
21 cm intensity mapping has emerged as a promising technique to map the 
large-scale structure of the Universe, at redshifts $z$ from 1 to 10.
Unfortunately, many of the key cross-correlations with the cosmic microwave 
background and photo-$z$ galaxies have been thought to be impossible due to 
the foreground contamination for radial modes with small wave numbers. 
In this paper, we apply tidal reconstruction to the simulated 21 cm fields
and recover the lost large-scale radial modes successfully.
We estimate the detectability of the cross-correlation signals and find they
can be detected at high significance with current 21 cm experiments.
The tidal field reconstruction method opens up a new set of possibilities to 
probe the Universe and is extremely valuable not only for 21 cm surveys
but also for cosmic microwave background and photometric-redshift observations.
\end{abstract}

\maketitle

\section{Introduction}
The current and future cosmic surveys aim to map a large fraction of the 
Universe with unprecedented precision by observing the large-scale structure 
(e.g., SDSS \cite{2017SDSS}, DES \cite{2016DES}, PFS \cite{2014PFS}, DESI 
\cite{2016DESI}, LSST \cite{2009LSST}, Euclid \cite{2012Euclid}) and the cosmic
microwave background (CMB) (e.g., Planck \cite{2016Planck}, SPT-3G \cite{2014SPT-3G}, Advanced ACTPol \cite{2016AdvACTPol}, CMB-S4 \cite{2016CMB-S4}). 
Precision measurement of cosmological parameters from the autocorrelations
of large-scale structure and CMB observations and the cross-correlations 
between different observations can improve constraints on the properties of 
dark energy, modifications to general relativity, neutrino masses, and 
primordial non-Gaussianities substantially.
In addition to these observation methods, 21 cm  intensity mapping has 
emerged as a powerful method to map the large-scale structure of the Universe 
\cite{2008IM,2008IM2,2010IM}. 
Instead of resolving millions of individual galaxies, the 21 cm intensity 
mapping technique measures the large-scale structure by detecting the aggregate 21 cm emission of neutral hydrogen from many galaxies in large voxels. 
The redshifted 21 cm emission line which provides the redshift information can 
be resolved exquisitely in the frequency domain.
Therefore, this allows radio telescopes to conduct rapid and efficient surveys
of large volumes of the Universe.
The ongoing and upcoming 21 cm surveys including CHIME \cite{2014CHIME}, HIRAX 
\cite{2016HIRAX}, Tianlai \cite{2015Tianlai}, BINGO \cite{2013BINGO}, FAST 
\cite{2016FAST}, MeerKAT \cite{2017MeerKAT}, and SKA \cite{2015SKA} can improve 
the baryon acoustic oscillations (BAO) measurements by observing a larger cosmic
volume at higher redshifts compared to current galaxy surveys.

The primary challenge for 21 cm intensity mapping experiments is the presence
of the astrophysical foregrounds from galactic and extra-galactic synchrotron 
emissions, which are three orders of magnitude brighter than the cosmological
21 cm signals. 
The synchrotron foregrounds are known to be spectrally smooth in the frequency
domain where the redshifted 21 cm signals from different redshifts fluctuate at 
different frequencies.
In principle, the foregrounds only impact the long wavelength density 
fluctuations along the line of sight, i.e., the modes with small $\kpa$ in 
Fourier space \cite{2003FG,2006FG}. 
However, the instrumental effects (e.g. spectral response, calibration, etc) 
further lead to an unsmooth foreground component, often referred as the 
foreground wedge at the low $\kpa$ and high $\kpe$ area in Fourier space 
\cite{2010Wedge,2012Wedge,2012Wedge2,2013Wedge,2015Wedge}.
The synchrotron foregrounds can be cleaned by exploiting their smooth spectral 
structure \cite{2011FG,2012FG,2014FG,2015FG}.
As demonstrated in Ref. \cite{2015FG}, the foregrounds can be cleaned well 
below the foreground wedge with the precise calibration of the instrument, 
leaving only $\kpa<0.02\ h\mr{Mpc}^{-1}$ modes unaccessible.

However, while there are not many Fourier modes at $\kpa<0.02\ h\mr{Mpc}^{-1}$, 
many other cosmological observations such as weak lensing, photometric-redshift
galaxies, and integrated Sachs-Wolf (ISW) effect can only probe these modes, i.e., the angular density fluctuations.
These observations involve a broad window function along the line of sight and
measure the projected modes, i.e., the modes with small $\kpa$, which are all
contaminated by the foreground emissions in 21 cm intensity mapping 
observations.
Therefore, the proposed cross-correlation of 21 cm intensity mapping with weak
gravitational lensing \cite{2015xLensing&Photoz,2016xLensing,2016xLensing2}, 
photo-$z$ galaxies \cite{2015xLensing&Photoz,2015xPhotoz,2015xPhotoz2,2017xPhotoz,2017xPhotoz&ISW,2017xPhotoz2}, and ISW effect \cite{2017xPhotoz&ISW} would 
be severely degraded in the presence of foregrounds.
Recovering the lost large-scale radial modes for cross-correlations is thus 
crucial in order to fully exploit the 21 cm intensity mapping experiments.
The cross-correlation measurements will benefit other observations as well,
since the cross-correlations are expected to be more robust to the systematics
than autocorrelations from individual experiments. 
It also enables the use of sample variance cancellation technique \cite{2009MT}
to measure cosmological parameters \cite{2015xPhotoz,2015xPhotoz2,2017xPhotoz}.

Recently a new method called cosmic tidal reconstruction has been developed
\cite{2012Tides,2016Tides}. 
The small-scale density fluctuations are significantly affected by the 
large-scale density field as a consequence of gravitational mode coupling.
The large-scale tidal shear field causes anisotropic distortions of the locally
measured small-scale matter power spectrum. 
Such local anisotropic tidal distortions can be exploited to reconstruct the 
large-scale tidal shear and hence density fields.
The reconstruction of gravitational tidal fields from local small-scale matter 
power spectrum is described by the same formulation as the reconstruction of 
gravitational lensing induced shear. 
As shown in Ref. \cite{2016Tides}, the density modes with small $\kpa$ 
and large $\kpe$ are well reconstructed, with cross-correlation coefficient
close to $1$ for reconstruction with the full dark matter density field.
These reconstructed density modes are exactly those lost in the foreground 
subtraction of 21 cm experiments. 
The tidal reconstruction technique enables the reconstruction of lost 21 cm 
radial modes, which provides important radial information essential for 
cross-correlating with the CMB and photometric observations.

In this paper, we apply cosmic tidal reconstruction to the simulated 21 cm 
field with low $\kpa$ foreground modes subtracted. The small $\kpa$ radial 
modes are recovered successfully after tidal reconstruction.
Then we cross-correlate the reconstructed 21 cm field with the simulated CMB 
lensing, photo-$z$ galaxy and ISW effect fields and estimate the detectability
of the cross-correlation signals with the current 21 cm experiments.
The tidal field reconstruction method provides us a new way to study the 
large-scale matter distribution in the Universe through cross-correlations 
and has profound implications for the current and future 21 cm experiments.

The paper is organized as follows.
In Sec. \ref{sec:rec}, we introduce the cosmic tidal reconstruction.
In Sec. \ref{sec:sim}, we apply tidal field reconstruction to the simulated 
foreground subtracted 21 cm density field and shows the reconstruction results.
Section \ref{sec:ccs} shows the cross-correlation signal recovered after
reconstruction and estimates the detectability with current 21 cm experiments.
We discuss further improvements and future applications in Sec. \ref{sec:dis}.

\section{Cosmic tidal reconstruction}
\label{sec:rec}
The large-scale density field can be reconstructed accurately from the 
anisotropic tidal distortions of the locally measured matter power spectrum
\cite{2012Tides,2016Tides}.
The basic idea of purely transverse tidal reconstruction has been proposed in 
Ref. \cite{2012Tides} and further expanded in Ref. \cite{2016Tides}.
In this section, we briefly discuss the physical idea and outline the 
operational procedure of the tidal field reconstruction. More details of this 
reconstruction method are presented in Ref. \cite{2016Tides}.

\subsection{Cosmic tides}

The evolution of small-scale density perturbations is modulated by long 
wavelength perturbations during nonlinear structure formation. 
The gravitational coupling of a long wavelength tidal field with small-scale
density fluctuations has been studied extensively \cite{2014Tides}.
The leading-order observable of a long wavelength density perturbation on 
small-scale density perturbations is described by the large-scale tidal field,
\bea    
t_{ij}=\Phi_{L,ij}-\delta_{ij}\Phi_{L,kk}/3,
\eea
where $\Phi_L$ is the long wavelength gravitational potential sourced by the 
long wavelength density perturbation $\delta_L$. Here $\Phi_{L,ij}$ denotes
partial derivatives of $\Phi_L$ to $x^i$ and $x^j$.
Note that we have projected out the trace of $\Phi_{L,ij}$, which corresponds 
to the local mean density.
Since the change of shape is more robust than the change of number density, we 
shall focus on the gravitational tidal shear, i.e., the traceless tidal field.
The locally observed matter power spectrum in the presence of the large-scale
tidal field $t_{ij}$ can be calculated using Lagrangian perturbation theory 
and is given by
\bea
\label{eq:pk}
P(\bmk,\tau)|_{t_{ij}}=P(k,\tau)+\hat{k}^i\hat{k}^jt^{(0)}_{ij}P(k,\tau)f(k,\tau),
\eea
where $\tau$ is the conformal time, $P(k,\tau)$ is the isotropic linear power 
spectrum, $\hat{\bmk}$ is the unit vector, the superscript $(0)$ denotes the 
initial time defined in perturbation calculation. 
The coupling of the large-scale tidal field to small-scale density fluctuations
is described by the tidal coupling coefficient
\bea
f(k,\tau)=2\alpha(\tau)-\beta(\tau)\frac{d\:\mr{ln}\:P(k,\tau)}{d\:\mr{ln}\:k},
\eea
where $\alpha(\tau)$ and $\beta(\tau)$ are integrals involving background cosmological parameters and can be computed numerically \cite{2014Tides, 2016Tides}.
The above result only includes the leading order effect of the coupling between
the large-scale tidal field and small-scale density fluctuations. 
In reality the density field is quite nonlinear and involves all higher order 
interactions. 
The reconstructed density field would be biased when the theoretical description
of the nonlinear coupling in the above equation is not accurate.
This problem can be addressed using the transfer function calibrated from 
simulations \cite{2016Tides}. 

The traceless tidal tensor $t_{ij}$ can be decomposed into five independently 
observable components ($\gamma_1$, $\gamma_2$, $\gamma_x$, $\gamma_y$, 
$\gamma_z$) \cite{2016Tides}.
We notice that the two transverse shear terms, 
\bea
\gamma_1=(\Phi_{L,11}-\Phi_{L,22})/2,\ \gamma_2=\Phi_{L,12},
\eea
which describe quadrupolar distortions in the tangential plane perpendicular 
to the line of sight, are less affected by peculiar velocities.
Thus, in the following computation we shall use them to perform reconstruction.
Once we have the tidal shear terms $\gamma_1$ and $\gamma_2$, the reconstructed
density field can be obtained by
\bea
\label{eq:rec}
\delta_r(\bmk)=\frac{Ak^2}{(k_1^2+k_2^2)^2}\bigg[(k_1^2-k_2^2)\gamma_1(\bmk)+2k_1k_2\gamma_2(\bmk)\bigg],
\eea
where $A$ is the normalization coefficient.
Since we only use two transverse tidal shear fields $\gamma_1$ and $\gamma_2$
for reconstruction, the change of the large-scale density field along the line 
of sight is inferred from the variations of $\gamma_1$ and $\gamma_2$ along 
the $z$ axis. 
The noise for the reconstructed density field is anisotropic in Fourier space.
The tidal reconstruction technique works best for modes in the high $\kpe$ 
and low $\kpa$ region, which cannot be obtained from 21 cm surveys directly 
but contribute substantially to observables from other cosmological observations
as discussed above. Cosmic tidal reconstruction provides a new possible way 
to recover the lost radial modes and to improve the cross-correlation signals.

\subsection{Reconstruction algorithm}
In this subsection, we describe the tidal reconstruction method used in the 
next section.

\subsubsection{Reducing nonlinearities}

The first step is to smooth the nonlinear density field with a Gaussian kernel, 
\bea
\delta_R(\bmk)=W_R(\bmk)\delta(\bmk),
\eea
where 
\bea
W_R(\bm{k})=\mr{exp}(-k^2R^2/2),
\eea
which filters out small-scale structures.
The perturbative description of tidal coupling in Eq. (\ref{eq:pk}) is not valid
in the strong non-Gaussian regions. 
We need to smooth small-scale nonlinear structures to reduce nonlinearities.
Here, we take $R=1.25\ \mr{Mpc}/h$, which is close to the optimal filter scale
as demonstrated in Refs. \cite{2012Tides,2016Tides}.  

The second step is to Gaussianize the smoothed density field by taking a 
logarithmic transform or mapping the density fluctuations into a Gaussian 
distribution according to their density values. 
We shall use the latter method since the tidal shear estimator we use is derived under the Gaussian assumption. Otherwise we can only use the limited number of density modes on large scales where the Gaussian assumption is valid, but the reconstruction will be degraded significantly.

\subsubsection{Estimating tidal shear fields}

The coupling of the large-scale tidal field and small-scale density fluctuations
leads to a local anisotropy of quadratic statistics.
The tidal shear fields can be reconstructed by applying quadratic estimators to
the Gaussianized density field $\delta_g$ as
\begin{eqnarray}
&&\hat{\gamma}_1(\bmx)=\big[\delta_g^{w_1}(\bmx)\delta_g^{w_1}(\bmx)+
\delta_g^{w_2}(\bmx)\delta_g^{w_2}(\bmx)\big]/2, \nonumber \\
&&\hat{\gamma}_2(\bmx)=\big[\delta_g^{w_1}(\bmx)\delta_g^{w_2}(\bmx)\big],
\end{eqnarray}
where 
\bea    
\delta_g^{w_1}(\bmk)=i\hat{k}_1w(\bmk)\delta_g(\bmk),\ 
\delta_g^{w_2}(\bmk)=i\hat{k}_2w(\bmk)\delta_g(\bmk),
\eea
and the filter is
\bea
w(\bmk)=\frac{\sqrt{P(\bmk)f(\bmk)}}{P_\mr{tot}(\bmk)},
\eea
Here, $P_\mr{tot}(\bmk)$ is the total power spectrum of
the 21 cm density field which includes both the signal and noise.
We find that the reconstruction performance is not sensitive to the exact shape of the filter.
The reconstruction of tidal shear fields is similar to the reconstruction of
lensing shear fields from 21 cm temperature fields.
The quadratic estimators presented above can be constructed using either 
the maximum likelihood method or the inverse variance weighting 
\cite{2008LensingEst,2010LensingEst,2012LensingEst}.

\subsubsection{Generating the density field}

After we get the tidal shear fields $\gamma_1$ and $\gamma_2$, the tidal 
reconstructed density field is given by Eq. (\ref{eq:rec}).
In general, the reconstructed field $\delta_r(\bmk)$ can be written as
\bea
\delta_r(\bmk)=C(\bmk)\delta(\bmk)+N(\bmk),
\eea
where $C(\bmk)=P_{\delta_r\delta}(\bmk)/P_{\delta}(\bmk)$, $\delta(\bmk)$ is 
the original  matter density field and $N(\bmk)$ includes the noises from
21 cm observation and from reconstruction. 
The factor $C(\bmk)$, often referred as the propagator, quantifies how much 
information of the original density distribution is reconstructed.
To get an unbiased measurement of the original density field, we can deconvolve 
the propagator $C(\bmk)$ from the reconstructed field,
\bea
\label{eq:delta}
\hat{\delta}_r=\delta_r(\bmk)/C(\bmk)=\delta(\bmk)+N(\bmk)/C(\bmk).
\eea
This factor can be computed by performing reconstruction with the simulated
observation mock data \cite{2017NR2,2017FM,2017NR3}. 
As we deconvolve the propagator, the reconstructed density field is unbiased. The cross-correlation coefficient quantifies the reconstruction noise.

\section{Implementation and results}
\label{sec:sim}

To test the performance of reconstruction, we run an ensemble of six $N$-body
simulations with the ${\tt CUBEP^3M}$ code \cite{2013CUBEP3M}. 
Each simulation involves $1024^3$ dark matter particles in a cubic box of side
length $1200\ \mr{Mpc}/h$. 
We use the snapshot at redshift $z=1$ and generate the dark matter density 
field on a $1024^3$ grid.
We could approximately use the dark matter density to represent the 21 cm 
source distribution, i.e., the neutral hydrogen.
This is a good approximation since the neutral hydrogen traces the total mass
distribution fairly well at low redshifts (see Refs. \cite{2017HI,2017Void}
for more discussions about the modeling of neutral hydrogen in the Universe).
However, the realistic neutral hydrogen density field should also include the fluctuation of neutral hydrogen fraction in the Universe, the redshift space distortion effect due to the peculiar velocity, etc. We plan to study these in future.

There are several noises for 21 cm experiments we need to consider to model
the observed 21 cm signal from intensity mapping observations, including the 
astrophysical foreground, the receiver noise, and the shortest baseline for 
inteferometers.

A detailed 21 cm foreground subtraction simulation is beyond the scope of this 
paper. Instead we simply use a high-pass filter along the line of sight,
\bea
W_{\mr{fs}}(\kpa)=1-e^{-\kpa^2R_\mr{fs}^2/2},
\eea
which removes the small $\kpa$ density modes, to simulate the loss of modes due
to foreground contamination. 
We use the two different foreground scales $R_\mr{fs}=60\ \mr{Mpc}/h$ and 
$15\ \mr{Mpc}/h$ in reconstruction, which give $W_{\mr{fs}}=0.5$ at 
$\kpa=0.02\ h\mr{Mpc}^{-1}$ and $0.08\ h\mr{Mpc}^{-1}$, respectively. 
The former is an optimal case, i.e., we only lose modes with 
$\kpa\lesssim0.02\ h\mr{Mpc}^{-1}$ \cite{2015FG}, while the latter is already 
achieved in the current 21 cm observations \cite{2013IM,2013IM2}.

For 21 cm observations around redshift $z=1$, the resolution of small-scale
structures is mainly determined by the thermal noise.
The thermal noise power $P_N$ is about $150-600\ (\mr{Mpc}/h)^3$ for a 
HIRAX-like interferometer, depending on the neutral hydrogen fraction and bias
\cite{2017Void}.
We assume the experimental noise to be zero above a cut off scale and infinity 
below this scale.
We choose it to be $k_N=0.6\ h\mr{Mpc}^{-1}$, which is about the scale 
where the thermal noise power dominates over the matter power spectrum.
The effect of the experimental noise can be modeled by applying a step function
\begin{displaymath}
\Theta(k_N-k)=\left\{
    \begin{array}{ll}
        1,& k\leq k_N  \\
        0,& k> k_N
    \end{array}\right.,
\end{displaymath}
to the dark matter density field from the simulation.

Most current 21 cm intensity mapping experiments are carried on interferometers.
The largest angular scale that can be probed is decided by the shortest 
baseline of the interferometer. 
We also use a step function
\begin{displaymath}
    \Theta(\ell-\ell_s)=\left\{
    \begin{array}{ll}
        1,& \ell\geq\ell_s  \\
        0,& \ell<\ell_s 
    \end{array}\right.,
\end{displaymath}
to model this effect, where the largest scale can be probed is $\ell_s=115$
or $\kpe=0.05\ h/\mr{Mpc}$ at redshift $z=1$.
This corresponds to a shortest baseline of $\sim7\ \mr{m}$.

In summary, the simulated 21 cm field from intensity mapping is given by 
\bea    
\delta_{\mr{IM}}(\bmk)=\delta(\bmk)W_{\mr{fs}}(\kpa)\Theta(k_N-k)\Theta(\ell-\ell_s),
\eea
where $\delta(\bmk)$ is the full density field from the simulation.
Note that $\ell$ is the angular wave number, defined as $\ell+1/2=k_\perp\chi(z)$.
We apply tidal reconstruction to the simulated 21 cm field and get the 
reconstructed density field defined in Eq. (\ref{eq:delta}) using the algorithm
described above.

\begin{figure}[tbp]
\begin{center}
\includegraphics[width=0.48\textwidth]{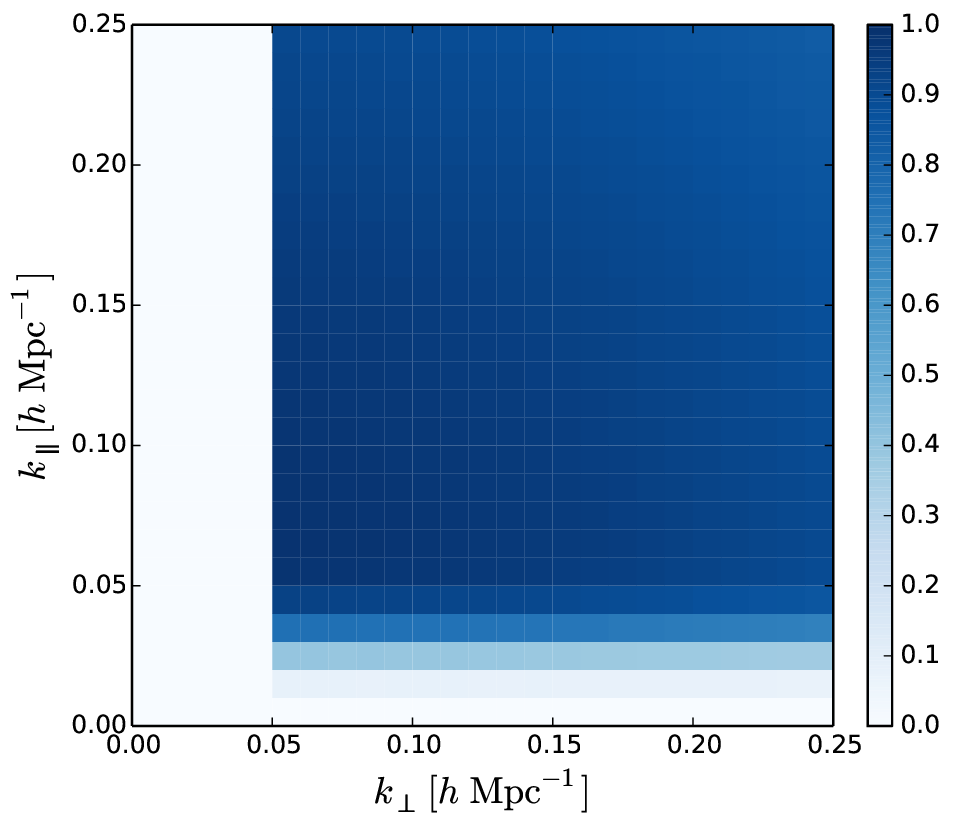}\\
\includegraphics[width=0.48\textwidth]{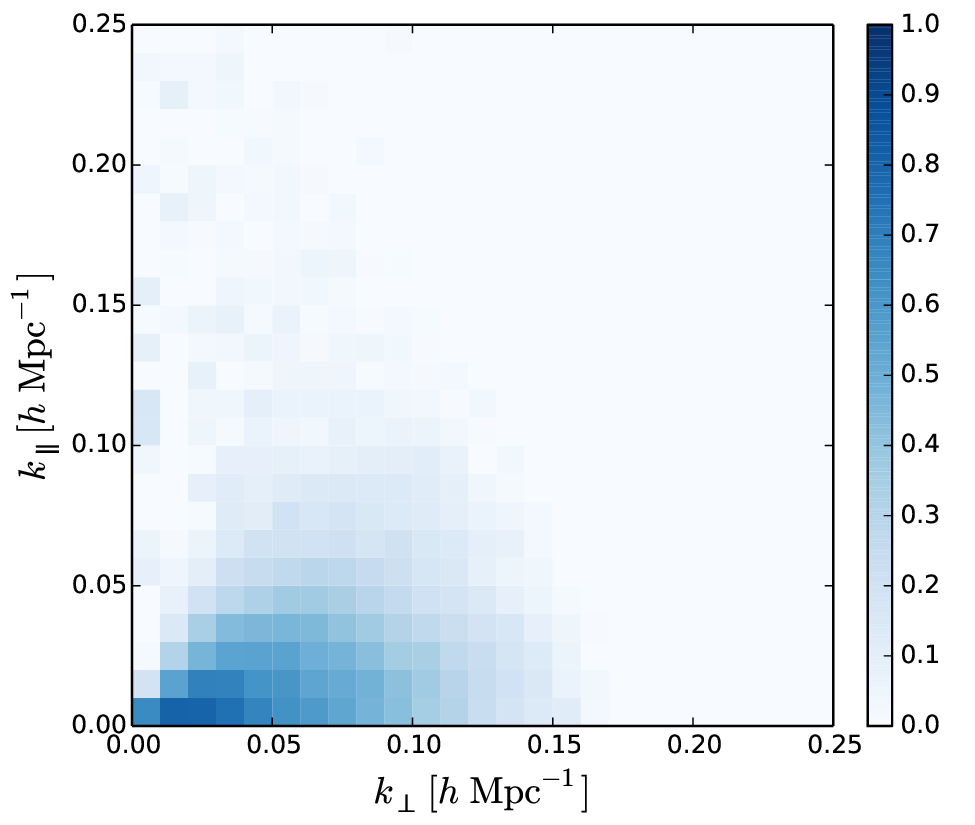}
\end{center}
\vspace{-0.7cm}
\caption{Top: The cross-correlation coefficient of the simulated 21 cm 
    intensity mapping field $\delta_\mr{IM}(\bmk)$ with the original full dark 
    matter density field $\delta(\bmk)$.
    Bottom: The cross-correlation coefficient of the reconstructed 
    density field $\hat{\delta}_r(\bmk)$ with the original full dark matter 
    density field $\delta(\bmk)$. 
    These results are for the foreground scale $R_\mr{fs}=60\ \mr{Mpc}/h$, 
    where modes with $\kpa\lesssim0.02\ h\mr{Mpc}^{-1}$ are subtracted.}
\label{fig:IM}
\end{figure}

\begin{figure}[tbp]
\begin{center}
\includegraphics[width=0.48\textwidth]{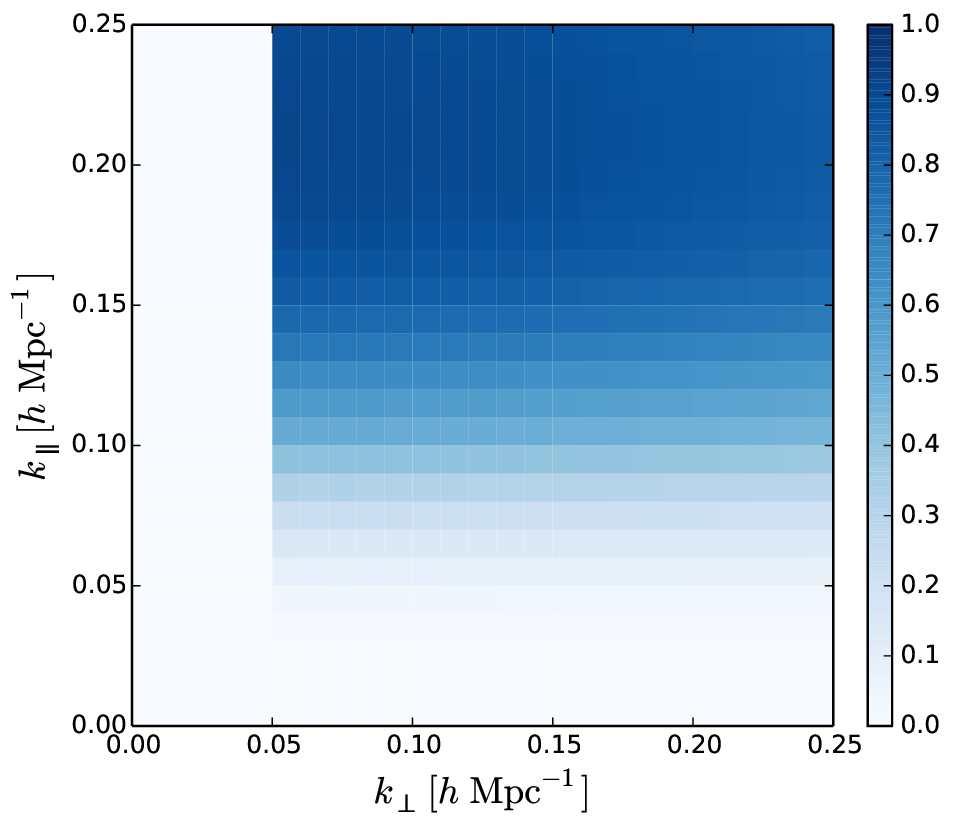}\\
\includegraphics[width=0.48\textwidth]{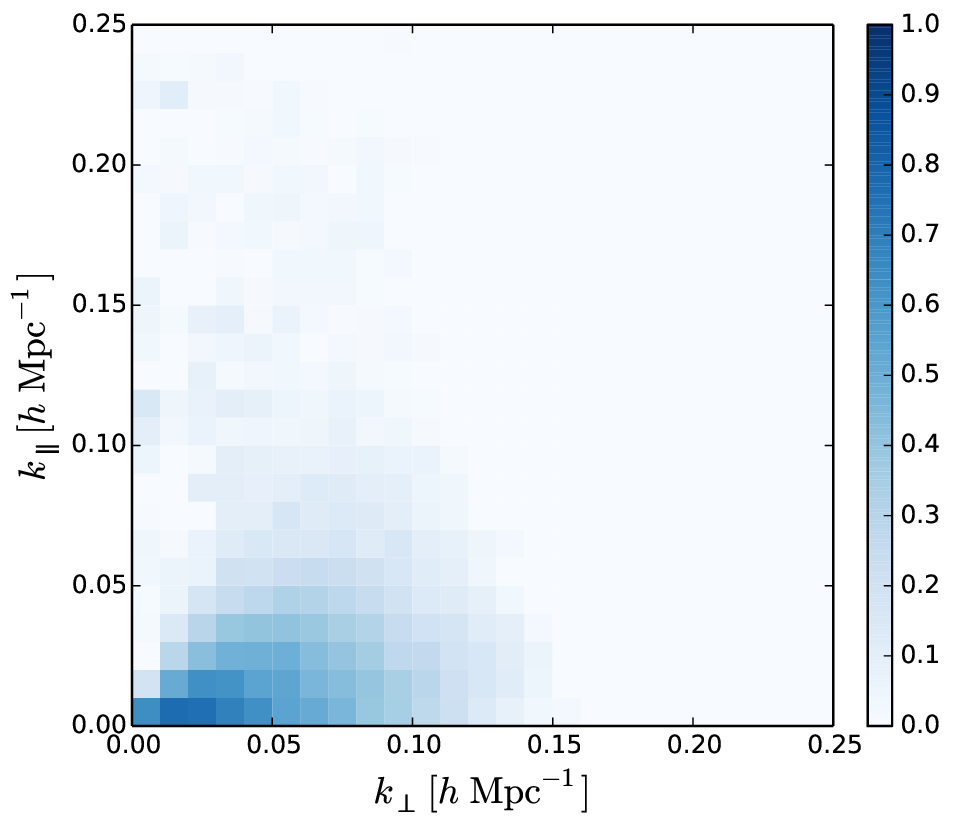}
\end{center}
\vspace{-0.7cm}
\caption{Same as Fig. \ref{fig:IM}, but for the foreground scale 
$R_\mr{fs}=15\ \mr{Mpc}/h$, where modes with $\kpa\lesssim0.08\ h\mr{Mpc}^{-1}$ 
are subtracted.}
\label{fig:IM2}
\end{figure}

Figure \ref{fig:IM} shows the two-dimensional cross-correlation coefficient of
the 21 cm intensity mapping field with the full dark matter density field. 
We also plot the cross-correlation coefficient of the reconstructed density
field with the full dark matter density field.
These results are for the foreground scale $R_\mr{fs}=60\ \mr{Mpc}/h$, i.e.,
$\kpa=0.02\ h\mr{Mpc}^{-1}$.
The lost large-scale radial modes are successfully recovered by tidal 
reconstruction.
Figure \ref{fig:IM2} shows the corresponding results for the foreground scale
$R_\mr{fs}=15\ \mr{Mpc}/h$, i.e., $\kpa=0.08\ h\mr{Mpc}^{-1}$.
We note that the loss of more large-scale radial modes does not degrade the 
performance of reconstruction significantly.
This is because the tidal reconstruction method uses small-scale structure to 
reconstruct the large-scale density field and the reconstruction performance 
mainly depends on the number of small-scale modes.

\begin{figure}[tbp]
\begin{center}
\includegraphics[width=0.48\textwidth]{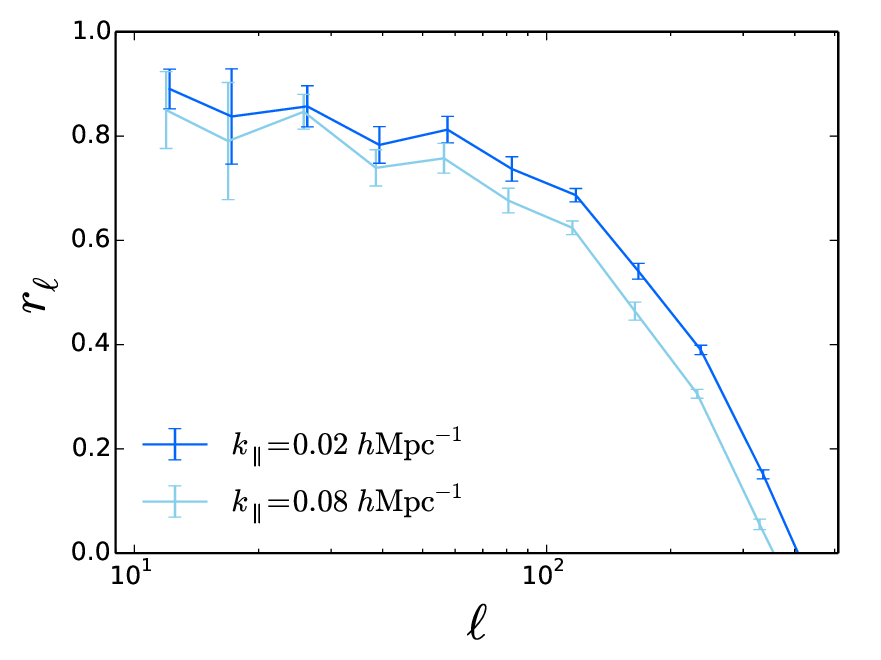}
\end{center}
\vspace{-0.7cm}
\caption{The cross-correlation coefficients of the projected full dark matter
density field with the projected reconstructed density fields for the 
foreground scales $\kpa=0.02\ h\mr{Mpc}^{-1}$ and $0.08\ h\mr{Mpc}^{-1}$.
The angular scale $\ell$ is related to the three-dimensional wave number $k$
through $\ell+1/2=k\chi(z=1)$, where $\chi(z=1)=2301\ \mr{Mpc}/h$ is the comoving
distance to redshift $z=1$. The cross-correlation coefficient is larger than 
$0.7$ and $0.65$ at scale $\ell\lesssim100$ for the small $\kpa$ and large 
$\kpa$ foreground scales, respectively.
The error bars are estimated using the bootstrap resampling method.}
\label{fig:cc}
\end{figure}

To clearly see how well the $\kpa\sim0$ modes relevant for cross-correlations 
are reconstructed, we compute the projected density field by averaging the 
three-dimensional density field along the line of sight, i.e., the $z$ axis
of the simulation box.
Figure \ref{fig:cc} shows the cross-correlation coefficients of the projected full dark matter density field with the projected reconstructed density fields for
the foreground scales $\kpa=0.02\ h\mr{Mpc}^{-1}$ and $0.08\ h\mr{Mpc}^{-1}$.
The angular scale $\ell$ is related to the three-dimensional wave number $k$
through $\ell+1/2=k\chi(z=1)$, where $\chi(z=1)=2301\ \mr{Mpc}/h$ is the comoving
distance to redshift $z=1$ \cite{2008Loverde}. The cross-correlation coefficient is larger than 
$0.7$ at scale $\ell\lesssim100$ for the small $\kpa$ foreground and larger 
than $0.65$ at scale $\ell\lesssim100$ for the large $\kpa$ foreground.
Therefore, the successful reconstruction of $\kpa\sim0$ modes makes the cross
correlation of  21 cm intensity mapping surveys with the CMB and photometric 
galaxy surveys possible.

\section{Cross-correlation signals}
\label{sec:ccs}

To estimate the detectability of the cross-correlation signals, we generate 
the CMB lensing convergence field, the angular galaxy distribution from 
photometric-redshift surveys, and the temperature fluctuation due to the ISW 
effect from the same simulation used for tidal reconstruction.
They are line of sight projections of the dark matter density field,
\bea    
\delta_i(\bm{\theta})=\int d\chi W_i(\chi)\delta(\chi\bm{\theta},\chi),
\eea
where $W_i(\chi)$ is the window function.
The angular cross-correlation power spectrum is then given by
\bea
\label{eq:xpk}
C_\ell^{ij}=\int d\chi\frac{W_i(\chi)W_j(\chi)}{\chi^2}P_\delta\bigg(k=\frac{\ell+1/2}{\chi},\chi\bigg).
\eea
When $i=j$, this formula gives the power spectrum for $\delta_i(\bm{\theta)}$.
The error for the cross-correlation signal is 
\begin{align}
\sigma(C_\ell^{ij})=\bigg[\frac{1}{(2\ell+1)f_\mr{sky}\Delta\ell}
    ((C_\ell^{ij})^2+\hat{C}_\ell^i
\hat{C}_\ell^j)\bigg]^{1/2},
\end{align}
where $\hat{C}$ includes the signal and the corresponding noise.
We set $\ell_\mr{min}=12$ and choose $f_\mr{sky}$ to be $0.25$ for CMB 
lensing and photo-$z$ galaxies, and $1$ for ISW effect. 
Notice that we use the projection of the reconstructed 21 cm field for cross
correlation.

\subsection{CMB lensing}

The lensing convergence field from CMB lensing reconstruction is a weighted 
projection of the dark matter density fluctuations along the line of sight 
to the last scattering surface,
\bea
\kappa(\bm{\theta})=\int_0^{\chi_s}d\chi 
W_\kappa(\chi)\delta(\chi\bm{\theta},\chi),
\eea
where the lensing kernel
\bea
W_\kappa(\chi)=\frac{3\Omega_{m0}H_0^2\chi(\chi_s-\chi)}{2a(\chi)\chi_s},
\eea
and $\chi_s=\chi(z_s=1090)$.
Because the CMB lensing kernel is very broad in redshift, we take its value at
redshift $z=1$ in the line of sight projection.
The noise for CMB lensing measurement is assumed to be the same as the Planck 
2015 results \cite{2016PlanckLensing}.

\begin{figure}[tbp]
\begin{center}
\includegraphics[width=0.48\textwidth]{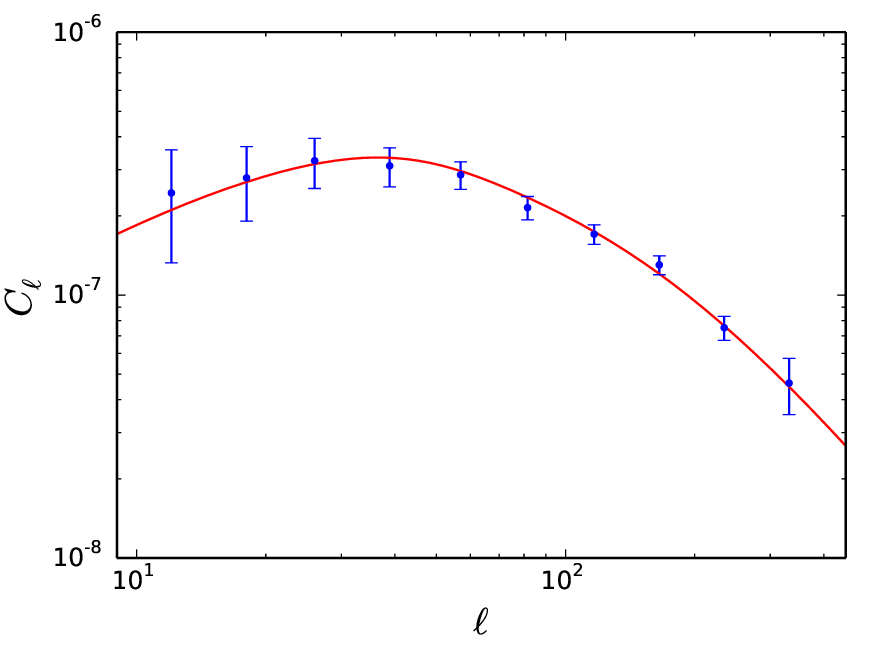}
\end{center}
\vspace{-0.7cm}
\caption{The cross-correlation signal of 21 cm intensity mapping and CMB 
    lensing. The solid line shows the theoretical cross power spectrum and the
    data points are measured from the simulations. 
    The error bars are for the foreground scale $R_\mr{fs}=60\ h/\mr{Mpc}$, 
    i.e., $\kpa=0.02\ h\mr{Mpc}^{-1}$.
}
\label{fig:xcc1}
\end{figure}

Figure \ref{fig:xcc1} shows the theoretical and measured cross power spectra. 
We also plot the error bars of the cross power spectrum for the $\kpa=0.02\ h\mr{Mpc}^{-1}$ foreground.
Since the error bars for the $\kpa=0.08\ h\mr{Mpc}^{-1}$ foreground are just 
slightly larger, we only plot the error bars for the small $\kpa$ foreground.
The total signal-to-noise ratio is $9.4$ and $8.0$ for the small and large
$\kpa$ foregrounds. 

\subsection{Photo-$z$ galaxies}

We calculate the projected galaxy density field at $z\sim 1$ with usual photo-$z$ bin width of $0.2$, i.e., $z_\mr{p}\in(0.9,1.1)$.
We adopt the galaxy distribution characterized by 
\bea
n(z)\propto z^{\alpha}\mr{exp}\big[-(z/z^{*})^\beta\big],
\eea
with $\alpha=2$, $z^*=0.5$, $\beta=1$ and assume the photometric-redshift
scatter $\mathcal{P}(z_\mr{p}|z)$ is perfectly known to be in a Gaussian form 
with photo-$z$ rms error $\sigma_z=0.05(1+z)$.
The angular galaxy distribution is given by
\bea
\delta_g(\bm{\theta})=\int_0^\infty dzW_\mr{p}(z)b(z)
\delta(\chi(z)\bm{\theta},\chi(z))\ ,
\eea
where the window function 
\bea
W_\mr{p}(z)\propto n(z)\int_{0.9}^{1.1} \mathcal{P}(z_\mr{p}|z) dz_\mr{p}\ 
\eea
with normalization $\int W_\mr{p}(z)dz=1$. 
We assume a linear galaxy bias $b(z)=1+0.84z$.
For photo-$z$ galaxies from LSST-like surveys, the shot noise is negligible
on degree scales. 

\begin{figure}[tbp]
\begin{center}
\includegraphics[width=0.48\textwidth]{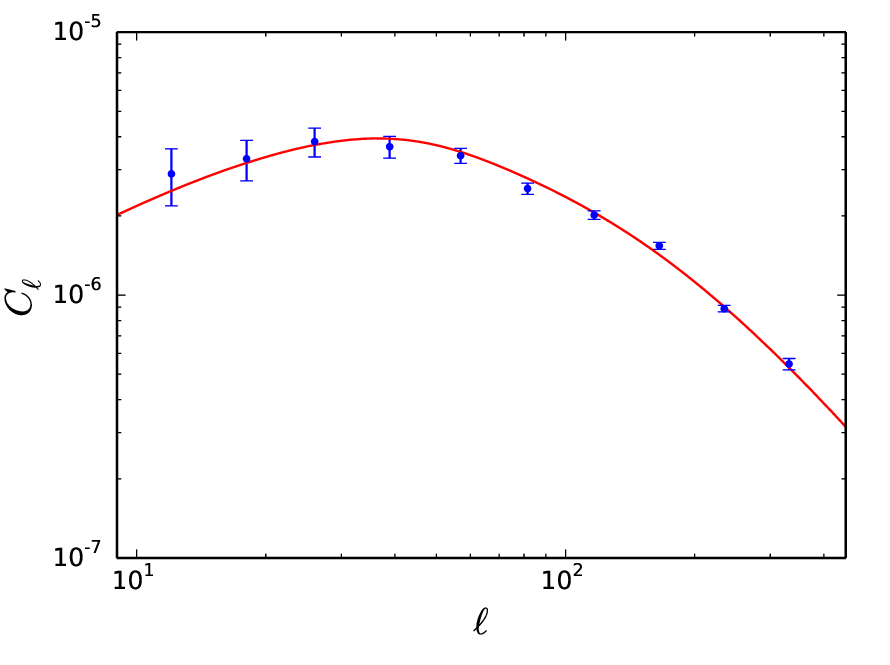}
\end{center}
\vspace{-0.7cm}
\caption{The cross-correlation signal of 21 cm intensity mapping and photo-$z$ 
    galaxies. The solid curve shows the theoretical cross power spectrum and 
    the data points are measured from the simulations. 
    The error bars are for the foreground scale $R_\mr{fs}=60\ h/\mr{Mpc}$, 
    i.e., $\kpa=0.02\ h\mr{Mpc}^{-1}$.
}
\label{fig:xcc2}
\end{figure}

Figure \ref{fig:xcc2} shows the theoretical and measured cross power spectra. 
We also plot the error bars for the foreground scale $\kpa=0.02\ h\mr{Mpc}^{-1}$.
The total signal-to-noise ratio is $24.3$ and $14.4$ for the small and large
$\kpa$ foregrounds, respectively.
The significant cross-correlation between reconstructed 21 cm field and photo-$z$ galaxies makes it possible to calibrate the redshift distribution of galaxies from imaging surveys using 21 cm intensity mapping surveys \cite{2017xPhotoz2}.

\subsection{ISW effect}

The fractional CMB temperature fluctuations induced by the 
ISW effect is given as 
\begin{eqnarray}
\bigg(\frac{\Delta T}{T}\bigg)_\mr{ISW}(\bm\theta)=-2\int_0^{\chi_s}d\chi
\frac{\partial\Phi(\chi\bm{\theta},\chi)}{\partial\chi}\ .
\end{eqnarray}
In Fourier space, approximating that the evolution of $\delta(\bm{k},t)$ with 
time is given by linear theory
$\dot\delta(\bm{k},t)=\dot D(t)\delta(\bm{k},t=0)$, we have
\begin{eqnarray}
\frac{\partial\Phi(\bm{k},\chi)}{\partial\chi}=-\frac{3\Omega_{m0}H_0^2}
{2a(\chi)}\frac{\partial\mr{ln}(D/a)}{\partial\chi}
\frac{\delta(\bm{k},\chi)}{k^2}\ ,
\end{eqnarray}
where $D$ is the linear growth function.
In our implementation, we approximate the time dependent factor as a 
constant across the simulation box.
For the ISW effect, the noise is just the large-scale CMB power $C_\ell^{TT}$. 

\begin{figure}[tbp]
\begin{center}
\includegraphics[width=0.48\textwidth]{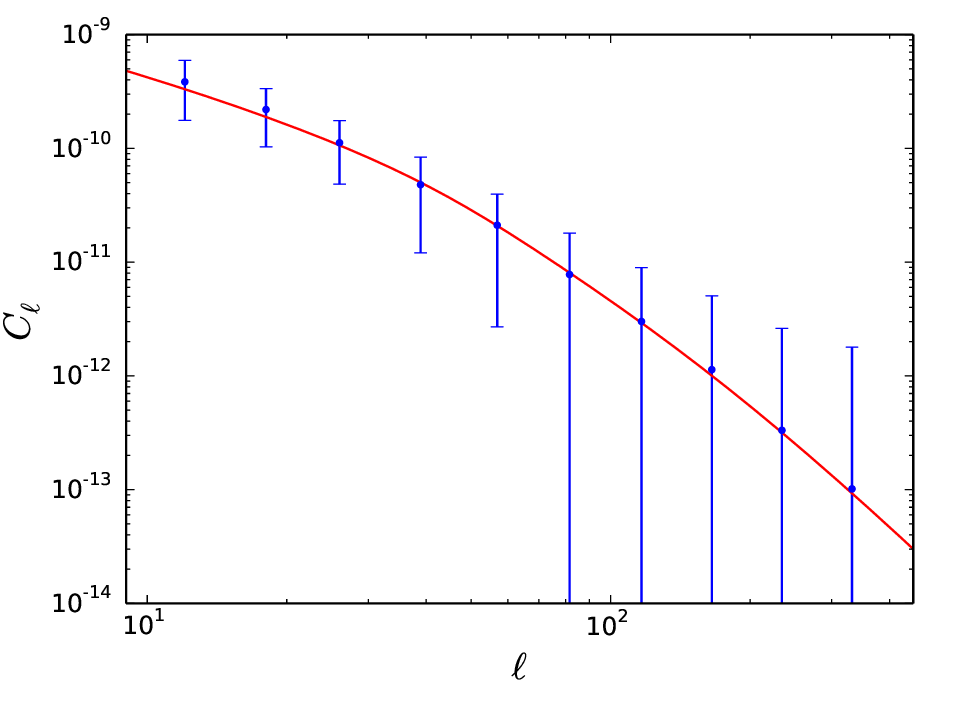}
\end{center}
\vspace{-0.7cm}
\caption{The cross-correlation signal of 21 cm intensity mapping and ISW effect.
    The solid curve shows the theoretical signal and the data points are 
    measured from the simulations. The error bars are for the foreground scale 
    $R_\mr{fs}=60\ h/\mr{Mpc}$, i.e., $\kpa=0.02\ h\mr{Mpc}^{-1}$.
}
\label{fig:xcc3}
\end{figure}

Figure \ref{fig:xcc3} shows the theoretical and measured cross power spectra.
We plot the error bars for the small $\kpa$ foreground.
The total signal-to-noise ratio is $3.1$ and $3.0$ for the small and large 
$\kpa$ foregrounds.
The redshift information from 21 cm intensity mapping allows us to constrain
the expansion history of the Universe as a function of redshift.
The detectability of ISW effect can be further improved by including CMB 
polarization data \cite{2011improveISW}.

\section{Discussion}
\label{sec:dis}

The detection significance presented here is for a 21 cm intensity mapping
survey of redshifts $0.8$--$1.4$, covering the quarter sky (full sky for 
ISW effect).
We use line of sight projections of the dark matter density field at 
redshift $z=1$ to approximate the observed cosmic fields. 
In reality, we need to consider the redshift evolution of density fluctuations
because of the relative wide redshift range.
The Limber approximation used to compute cross power spectrum is not accurate 
on very large scales.
Since angular power spectrum is larger when computed using exact integration 
than with the Limber approximation \cite{2017Limber,2017Limber2}, the detection
significance should not be degraded by this approximation.
On extremely large scales, the relativistic effects should also be included 
when predicting the angular power spectrum \cite{2013GRSignal,2017GRSignal}.
As there are not many modes measurable on extremely large scales, the 
realtivistic effects should not affect the results much.
However, to be conservative, we still use only the $\ell>12$ angular modes to 
estimate the detection significance.

The tidal shear estimators adopted here are optimal only for the Gaussian field
and in the long wavelength limit \cite{2016Tides,2008LensingEst,2012LensingEst}.
The results can be improved by constructing optimal estimators for non-Gaussian 
fields as have done in 21 cm lensing \cite{2010LensingEst}.
The correlation coefficient drops quickly towards small scales.
This is because there is not enough small-scale modes in 21 cm intensity mapping
surveys and the estimators are not optimal in the equilateral configuration.
The long wavelength optimal estimators relies on the number of small-scale
modes; the performance would be better with more small-scale structures. 
The tidal reconstruction can still be improved by developing new algorithms to
deal with the nonlinear coupling beyond the squeezed configuration.

The BAO reconstruction technique has been shown to be still useful in 21 cm 
surveys \cite{2016BAO,2016Combine,2017BAO}.
While there are not many modes with small $\kpa$ lost due to the foreground, the
differential motions which smear the BAO peak are substantially contributed by 
large-scale modes with $k\lesssim0.1\ h\mr{Mpc}^{-1}$.
Performing nonlinear reconstruction also needs these modes to estimate the 
large-scale linear displacement \cite{2017NR2,2016NR,2017NR}.
Cosmic tidal reconstruction compensates the foreground wedge at small $\kpa$
and large $\kpe$ and hence can improve the BAO measurements from 21 cm surveys 
\cite{2016BAO,2016Combine}. 
These recovered foreground modes can also improve the efficiency of the void 
finder with interferometric 21 cm experiments \cite{2017Void}.
In addition to the cross-correlations explored here, the tidal reconstruction 
method also works for the kinematic Sunyaev-Zel'dovich effect which we leave 
for future work.

All cosmological 21 cm experiments share the same foreground problem, no matter
low redshift BAO experiment or high redshift epoch of reionization observation.
Therefore, the tidal reconstruction method is also important for high redshift 
experiment such as measuring the cross-correlation of 21 cm with the kinematic
Sunyaev-Zel'dovich effect from the epoch of reionization \cite{2016kSZ}.

\section*{Acknowledgements}
We thank Alex van Engelen, Marcelo Alvarez, Philippe Berger, 
Yi-Chao Li, Shifan Zuo, Wen-Xiao Xu, and Tian-Xiang Mao for helpful discussions.
We acknowledge the support of the Chinese Ministry of Science and Technology
under Grant No. 2016YFE0100300, the National Natural Science Foundation of 
China under Grants No. 11633004, No. 11373030, No. 11773048 and No. 11403071, 
CAS Grant No. QYZDJ-SSW-SLH017, and NSERC.
The simulations are performed on the BGQ supercomputer at the SciNet HPC 
Consortium. SciNet is funded by the following: the Canada Foundation for 
Innovation under 
the auspices of Compute Canada, the Government of Ontario, Ontario Research 
Fund---Research Excellence, and the University of Toronto.
The Dunlap Institute is funded through an endowment established by the David Dunlap family and the University of Toronto.
Research at the Perimeter Institute is supported by the Government of Canada 
through Industry Canada and by the Province of Ontario through the Ministry of 
Research and Innovation.

\bibliographystyle{apsrev}
\bibliography{cro}

\end{document}